\documentclass{phbauth}
\usepackage{graphicx}
\begin{document}

\begin{frontmatter}

\title{Contact phenomena in carbon nanotubes}

\author[address1,address2]{Arkadi A. Odintsov\thanksref{thank1}},
\author[address1,address3]{Yasuhiro Tokura},

\address[address1]{Department of Applied Physics and DIMES, 
Delft University of Technology,
2628 CJ Delft, The Netherlands}
\address[address2]{Nuclear Physics Institute, Moscow State University, 
Moscow 119899 GSP, Russia}
\address[address3]{NTT Basic Research Laboratories, Morinosato-Wakamiya, 
Atsugi-shi 243-01, Japan}

\thanks[thank1]{Corresponding author.
E-mail: odintsov@duttntn.tn. tudelft.nl}

\begin{abstract}
Poor screening of the long-range Coulomb interaction 
in one-dimensional carbon nanotubes 
results in a peculiar picture of contact phenomena.
Being brought to a contact with a metal, conducting
nanotube accumulates electric charge whose density decays  
slowly with the distance from the contact.
This should be contrasted to a conventional metal-metal contact
where the charge density decreases exponentially at atomic distances.
Implications for experiments  are discussed.
\end{abstract}

\end{frontmatter}

The contacts to metallic electrodes play a major role in the study of
electron transport in carbon nanotubes. The difference in the work functions
of a metal (Au,Pt) and graphite results in the charge transfer between the
nanotube and electrodes, which shifts the Fermi level of the nanotube
downward from half-filling. Corresponding shift of the Fermi vector $k_{F}$
away from the band-crossing point $K=2\pi /3a$ has been detected
experimentally in the pattern of standing electron waves in SWNTs on a gold
substrate \cite{Venema}.

In conventional metal-metal contacts the charge density induced by a
mismatch of the work functions decays within the Thomas-Fermi screening
distance ($\sim 0.5$ \AA ) away from the interface. Despite conducting SWNTs
have metallic electronic spectrum, the Coulomb screening is expected to be
substantially less effective due to reduced dimensionality of the system.
The analysis of contact phenomena in this case is the goal of our work.

For simplicity we assume that conducting electrons in SWNT are confined to
the surface of a cylinder of radius $R.$ The electrostatic potential $\Phi
(R,z)\sim \Phi _{q}e^{iqz}$ at the surface can be found from the solution of
the Poisson equation, 
\begin{equation}
\Phi _{q}=u_{q}\rho _{q},  \label{Eq_Phi}
\end{equation}
where $\rho _{q}e^{iqz}$ is a linear density of charge, $%
u_{q}=2I_{0}(qR)K_{0}(qR)/\kappa $ is a bare Coulomb interaction and $\kappa 
$ is an effective dielectric constant of the media. The charge density 
\begin{equation}
\rho _{q}=-e\nu _{q}W_{q}  \label{Eq_rho}
\end{equation}
is related to the total potential $W_{q}=V_{q}+e\Phi _{q}$ ($e>0$) induced
by the external potential $V_{q}$ \cite{Ziman}. 
\begin{figure}[t]
\par
\begin{center}
\leavevmode
\includegraphics[width=0.75\linewidth]{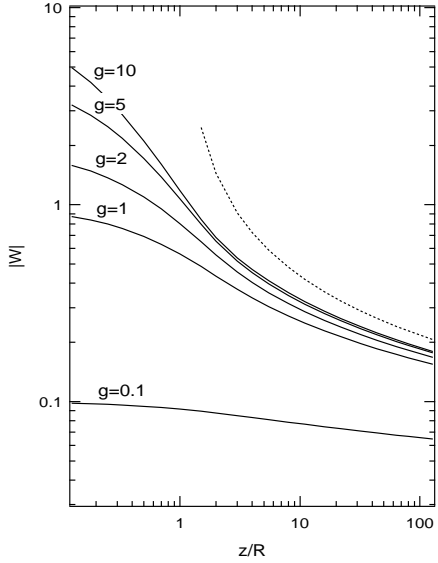}
\end{center}
\caption{The deviation $|W|$ of the Fermi level from half-filling for edge
contacted SWNT in units of $V/g$. Dashed line corresponds to Eq.~\ref
{rho_edge}.}
\label{figurename}
\end{figure}
In the long wave length limit ($q\ll K$) $\nu _{q}$ is given by the
single-particle density of states in SWNT, $\nu _{q}=\nu =4/(\pi \hbar
v_{F}) $. In this case the total potential $W(z)=-\rho (z)/e\nu $
corresponds to the deviation of the local Fermi level from half-filling.
Eqs. \ref{Eq_Phi}, \ref{Eq_rho} determine the response of the system $%
W_{q}=V_{q}/\varepsilon _{q}$, $\varepsilon _{q}=1+e^{2}\nu _{q}u_{q}$ to
the external potential $V_{q}$, which will be studied for three situations.

(i) Edge contact. Consider SWNT aligned along $z$ axis ($z>0$) and
contacting the plane $xy$ of a metallic electrode ($z<0$) at $z=0$. The
difference $V=\varphi _{M}-\varphi _{N}$ of the work functions of a metal ($%
\varphi _{M}$) and a nanotube ($\varphi _{N}$) plays a role of the external
potential which induces charge at the nanotube. In order to take the image
charges into account we introduce $\rho (-z)=-\rho (z)$ ($z>0$) and modify
the external potential accordingly, $V(z)=-Vsign(z)$. As a result we obtain, 
\begin{equation}
W(z)=-\frac{V}{g}\frac{1}{\ln (z/R)},  \label{rho_edge}
\end{equation}
for $z\gg R$ and $g\equiv 2e^{2}\nu /\kappa \gg 1$ ($2e^{2}\nu \approx 7.4$%
). Numerical results for moderate values of the interaction are presented in
Fig. 1.

(ii) Lateral point contact. We consider SWNT contacted by a small metallic
electrode at $z=0$. The external potential can be approximated by a delta
function $V(z)=V_{e}\delta (z)$ whose amplitude depends on the microscopic
details of the contact ($V_{e}\propto V$). The position of the Fermi level
at $|z|\gg R$ is given by, 
\begin{equation}
W(z)=-\frac{V_{e}}{2g}\frac{1}{|z|\ln ^{2}(|z|/R)},  \label{rho_point}
\end{equation}
for $g\gg 1$. The same relation with $-V_{e}/2g\rightarrow e^{2}/\kappa
g^{2} $ describes screened Coulomb interaction in isolated nanotube. Despite
the screened interaction is by a factor $g^{2}\gg 1$ weaker than unscreened
one, it decays only marginally faster.

(iii) Lateral long contact. Finally we treat SWNT on the surface of a metal.
Assuming that the charge is distributed uniformly around the circumference
of the cylinder (SWNT) and that the distance $d$\ between the cylinder and
the metal surface is small $d\ll R,$ we modify Eq. \ref{Eq_Phi} as follows, $%
\Phi =2\rho d/R.$ The ''external potential'' of the nanotube is fixed at $-V$
(cf. (i)). Using Eq. \ref{Eq_rho} we obtain $W=V/\varepsilon _{l}$ with $%
\varepsilon _{l}=1+2e^{2}\nu d/R$. For a nanotube on gold $V\simeq 0.8$ eV 
\cite{Venema}. Loosely estimating $d=a\simeq 2.46$ \AA\ we obtain the
deviation of the Fermi level from half-filling, $W\simeq 0.22$ eV for
(10,10) SWNT. The agreement with the experimental value $W_{\exp }$ $\simeq
0.3$ eV \cite{Venema} is reasonable, taking into account simplicity of our
model.

We conjecture that the shift of the Fermi level from half-filling can be
detected in the pattern of Friedel oscillations from impurities or ends of a
nanotube.

\begin{ack}
We thank G.E.W. Bauer, C. Dekker, and 
Yu.V. Nazarov for stimulating discussions. The
financial support of KNAW is gratefully acknowledged.
\end{ack}

\end{document}